\documentstyle[11pt]{article}
\begin{document}
\title{{\bf Simulations of Quantum Turing Machines by Quantum Multi-Stack
Machines\\ }}
\author{Daowen Qiu\thanks{
Email-address: issqdw@zsu.edu.cn (D. Qiu).}\\
{\footnotesize Department of Computer Science, Zhongshan
University, Guangzhou, 510275, P.R. China}\\}
\date{  }
\maketitle

\par
\vskip 1mm {\bf Abstract:} \hskip 5mm As was well known, in
classical computation, Turing machines, circuits, multi-stack
machines, and multi-counter machines are equivalent, that is, they
can simulate each other in polynomial time. In quantum
computation, Yao [11] first proved that for any quantum Turing
machines $M$, there exists quantum Boolean circuit
$(n,t)$-simulating $M$, where $n$ denotes the length of input
strings, and $t$ is the number of move steps before machine
stopping. However, the simulations of quantum Turing machines by
quantum multi-stack machines and quantum multi-counter machines
have not been considered, and quantum multi-stack machines have
not been established, either. Though quantum counter machines were
dealt with by Kravtsev [6] and Yamasaki {\it et al.} [10], in
which the machines count with $0,\pm 1$ only, we sense that it is
difficult to simulate quantum Turing machines in terms of this
fashion of quantum computing devices, and we therefore prove that
the quantum multi-counter machines allowed to count with $0,\pm
1,\pm 2,\ldots,\pm n$ for some $n>1$ can efficiently simulate
quantum Turing machines.

So, our mail goals are to establish quantum multi-stack machines
and quantum multi-counter machines with counts $0,\pm 1,\pm
2,\ldots,\pm n$ and $n>1$, and particularly to simulate quantum
Turing machines by these quantum computing devices. The major
technical contributions of this article are stated as follows:

(i) We define {\it quantum multi-stack machines} (abbr. QMSMs) by
generalizing a kind of {\it quantum pushdown automata} (abbr.
QPDAs) from one-stack to multi-stack, and the {\it
well-formedness} (abbr. W-F) conditions for characterizing the
unitary evolution of the QMSMs are presented.

(ii) By means of QMSMs we define {\it quantum multi-counter
machines} (abbr. QMCMs) whose state transition functions are
different from the {\it quantum counter automata} (abbr. QCAs) in
the literature; as well, the W-F conditions are given for these
defined QMCMs.

(iii) To simulate {\it quantum Turing machines} (abbr. QTMs), we
deal with a number of simulations between QMCMs with different
counters and different counts. Therefore, we show that any QMCM
allowed to count with $\pm n$ for $n>1$ can be simulated by
another QMCM that counts with $0,\pm 1$ only.

(iv) In particular, we demonstrate the efficient simulations of
QTMs in terms of QMSMs, and the simulation of QMCMs by QMSMs with
the same time complexity. Therefore, we show that QTMs can be
simulated by QMSMs as well. To conclude, a number of issues are
proposed for further considerations.


\section*{1. Introduction}

\subsection*{{\it 1.1. Motivation and purpose}}

Quantum computing is an intriguing and promising research field,
which touches on quantum physics, computer science, and
mathematics [4]. To a certain extent, this intensive attention
given by the research community originated from Shor's findings of
quantum algorithms for factoring prime integers in polynomial time
and Grover's algorithm for searching through a database which
could also be sped up on a quantum computer [4].

Let us briefly recall the work of pioneers in this area. (Due to
limited space, the detailed background and related references are
referred to [4].) In 1980, Benioff first considered that the
computing devices in terms of the principles of quantum mechanics
could be at least as powerful as classical computers. Then Feynman
pointed out that there appears to be no efficient way of
simulating a quantum mechanical system on a classical computer,
and suggested that a computer based on quantum physical principles
might be able to carry out the simulation efficiently. In 1985
Deutsch re-examined the Church-Turing Principle and defined QTMs.
Subsequently, Deutsch considered quantum network models.

Quantum computation from a complexity theoretical viewpoint was
studied systematically by Bernstein and Vazirani [1] and they
described an efficient universal QTM that can simulate a large
class of QTMs. Notably, in 1993 Yao [11] demonstrated the
equivalence between QTMs and quantum circuits. More exactly, Yao
[11] showed that any given QTM, there exists a quantum Boolean
circuit $(n,t)$-simulating this QTM with polynomial time slowdown,
where $n$ denotes the length of input strings, and $t$ is the
number of move steps before machine stopping.

In the theory of classical computation [5], both 2-stack machines,
as a generalization of pushdown automata, and 2-counter machines
can efficiently simulate Turing machines [7,2,5]. However, as the
authors are aware, the simulations of QTMs in terms of QMSMs and
QMCMs still have not been considered. Since Turing machines,
circuits, multi-stack machines, and multi-counter machines are
equivalent in classical computation, we naturally hope to clarify
their computing power in quantum computers. Therefore, our focuses
in this article are to introduce QMSMs and QMCMs that are somewhat
different from the QCAs in the literature [6,10], and
particularly, to simulate QTMs by virtue of these two quantum
computing devices.

Indeed, in quantum computing devices, the unitarity of evolution
operators is generally characterized by the W-F conditions of the
local transition function of the quantum models under
consideration. Bernstein and Vazirani [1] gave the W-F conditions
for the QTMs whose read/write heads are not allowed to be
stationary in each move. In QTMs whose read/write heads are
allowed to be stationary (called {\it generalized QTMs}, as in
[1]), the first sufficient conditions for preserving the unitarity
of time evolution were given by Hirvensalo, and then Yamasaki [12]
gave the simple W-F conditions for multiple-tape
stationary-head-move QTMs, and Ozawa and Nishimura further
presented the W-F conditions for the general QTMs. For the
details, see [4, p. 173]. Golovkins [3] defined a kind of QPDAs
and gave the corresponding W-F conditions; Yamasaki {\it et. al.}
[10] defined quantum 2-counter automata and presented the
corresponding W-F conditions, as well.

We see that those aforementioned W-F conditions given by these
authors for corresponding quantum computing devices are quite
complicated. Therefore, based on the QPDAs proposed in [9] where
QPDAs in [9] and [8] are shown to be equivalent, we would like to
define QMSMs that generalize the QPDAs in [9], and further define
QMCMs. As well, we will give the W-F conditions for these defined
devices. Notably, these W-F conditions are more succinct than
those mentioned above. In particular, motivated by Yao's work [11]
concerning the $(n,t)$-simulations of QTMs by quantum circuits, we
will use QMSMs to $(n,t)$-simulate QTMs, where $n$ denotes that
the length of input strings are not beyond $n$, and $t$ represents
that the number of move steps of QTMs (time complexity) is not
bigger that $t$ for those input strings.

\subsection*{{\it 1.2. Main results}}

According to the above analysis, we state the main contributions
in this article. In Section 2, we define QMSMs by generalizing
QPDAs in [9] from one-stack to muti-stack, and present the
corresponding W-F conditions (Theorem 1) for the defined quantum
devices.

In Section 3, by means of QMSMs we define QMCMs that are somewhat
different from the QCAs by Kravtsev [6] and Yamasaki {\it et al.}
[10]; as well, the W-F conditions (Theorem 2) are given for the
defined QMCMs. It is worth indicating that the state transition
functions in QCAs defined by Kravtsev [6] and Yamasaki {\it et
al.} [10] have local property, since they are defined on $Q\times
\{0,1\}\times (\Sigma\cup\{\#,\$\})\times Q\times \{0,1\}$, but
their W-F conditions are quite complicated, while in QMCMs defined
in this article, the state transition functions are on $Q\times
{\bf N}^{k}\times (\Sigma\cup\{\#,\$\})\times Q\times {\bf
N}^{k}$, and consequently, the corresponding W-F conditions are
more succinct (see Theorem 2).

To simulate QTMs, we deal with a number of properties regarding
simulations between QMCMs with different counters and different
counts (Lemmas 1 and 2). We show that QMCMs allowed to count with
$0,\pm 1,\pm 2,\ldots,\pm n$ can be simulated by QMCMs that are
able to count with $0,\pm 1$ only but need more counters.

In particular, in Section 4, we present the simulations of QTMs in
terms of QMCMs with polynomial time slowdown. More specifically,
we prove that any QTM $M_{1}$, there exists QMSM $M_{2}$
$(n,t)$-simulating $M_{1}$, where $n$ denotes the length of input
strings not bigger than $n$, and $t$ represents that the number of
move steps of QTMs is not bigger that $t$ for those input strings.
Also, we show that QMCMs can be simulated by QMSMs with the same
time complexity, and by this result it then follows the efficient
simulations of QTMs by QMSMs.

Due to the page limit, these detailed proofs of Theorems and
Lemmas are put in Appendices. In this extended abstract, notations
will be explained when they first appear.

\section*{2. Quantum multi-stack machines}

QPDAs were considered by the authors in [8,3,9]. Here we will
define quantum $k$-stack machines by generalizing the QPDAs in [9]
from one stack to $k$ stacks.

{\bf Definition 1.} A quasi-quantum two-stack machine is defined
as $M=(Q,\Sigma,\Gamma,\delta,Z_{0},q_{0},q_{a},q_{r})$ where $Q$
is the set of states, $\Sigma$ is the input alphabet, $\Gamma$ is
the stack alphabet, $Z_{0}\in\Gamma$ denotes the most bottom
symbol that is not allowed to be popped, $q_{0}\in Q$ is the
initial state, and $q_{a},q_{r}\in Q$ are respectively the
accepting and rejecting states, and transition function $\delta$
is defined as follows:
\[
\delta:
Q\times\Gamma^{*}\times\Gamma^{*}\times(\Sigma\cup\{\#,\$\})\times
Q\times\Gamma^{*}\times\Gamma^{*}\rightarrow {\bf C}
\]
where $\Gamma^{*}$ denotes the set of all strings over $\Gamma$,
and
$\delta(q,\gamma_{1},\gamma_{2},\sigma,q^{'},\gamma_{1}^{'},\gamma_{2}^{'})\not=0$
if and only if

(i) $\gamma_{1}=\gamma_{1}^{'}$ or $X\gamma_{1}=\gamma_{1}^{'}$ or
$\gamma_{1}=X\gamma_{1}^{'}$ for some $X\in\Gamma\backslash
\{Z_{0}\}$; and

(ii) $\gamma_{2}=\gamma_{2}^{'}$ or $Y\gamma_{2}=\gamma_{2}^{'}$
or $\gamma_{2}=Y\gamma_{2}^{'}$ for some $Y\in\Gamma\backslash
\{Z_{0}\}$.

\noindent In addition, if $\gamma_{1}=Z_{0}$, then
$\gamma_{1}^{'}=Z_{0}$ or $\gamma_{1}^{'}=ZZ_{0}$ for some $Z\in
\Gamma$ with $Z\not=Z_{0}$; similar restriction is imposed on
$\gamma_{2}$.

A configuration of the machine is described by
$|q\rangle|\gamma_{1}\rangle|\gamma_{2}\rangle$, where $q$ is the
current control state, $\gamma_{1}$ and $\gamma_{2}$ represent the
current strings of stack symbols in two stacks, respectively, and
we identify the leftmost symbol of $\gamma_{i}$ with the top stack
symbol of stack $i$ for $i=1,2$. Therefore, the rightmost symbol
of $\gamma_{i}$ is $Z_{0}$. We denote by $C_{M}$ as the set of all
configurations of $M$, that is,
\[
C_{M}=\{|q\rangle|\gamma_{1}Z_{0}\rangle|\gamma_{2}Z_{0}\rangle:
q\in Q, \gamma_{i}\in \Gamma^{*},i=1,2\}.
\]
Let $H_{X}$ represent the Hilbert space whose orthonormal basis is
the set $X$, that is, $H_{X}=l_{2}(X)$. Therefore, $H_{Q}\otimes
H_{\Gamma^{*}}\otimes H_{\Gamma^{*}}$ is a Hilbert space whose
orthonormal basis is $C_{M}$, that is, $l_{2}(C_{M})=H_{Q}\otimes
H_{\Gamma^{*}}\otimes H_{\Gamma^{*}}$. In addition, we assume that
there are endmarkers $\#$ and $\$$ representing the leftmost and
rightmost symbols for any input string $x\in\Sigma^{*}$.
Therefore, any input string $x\in\Sigma^{*}$ is put on the input
tape in the form of $\#x\$$, and the read head of $M$ begins with
$\#$ and ends after reading $\$$.

Intuitively,
$\delta(q,\gamma_{1},\gamma_{2},\sigma,q^{'},\gamma_{1}^{'},\gamma_{2}^{'})$
denotes the amplitude of the machine evolving into configuration
$|q^{'}\rangle|\gamma_{1}^{'}\rangle|\gamma_{2}^{'}\rangle$ from
the current one $|q\rangle|\gamma_{1}\rangle|\gamma_{2}\rangle$
after reading input $\sigma$.

For any $\sigma\in\Sigma\cup\{\#,\$\}$ we defined the time
evolution operators $U_{\sigma}$ and $U_{\sigma}^{'}$ from
$H_{Q}\otimes H_{\Gamma^{*}}\otimes H_{\Gamma^{*}}$ to
$H_{Q}\otimes H_{\Gamma^{*}}\otimes H_{\Gamma^{*}}$ as follows:
\begin{equation}
U_{\sigma}(|q\rangle|\gamma_{1}\rangle|\gamma_{2}\rangle)=
\sum_{q^{'},\gamma_{1}^{'},\gamma_{2}^{'}}\delta(q,\gamma_{1},\gamma_{2},\sigma,q^{'},\gamma_{1}^{'},\gamma_{2}^{'})
|q^{'}\rangle|\gamma_{1}^{'}\rangle|\gamma_{2}^{'}\rangle,
\end{equation}
\begin{equation}
U_{\sigma}^{'}(|q\rangle|\gamma_{1}\rangle|\gamma_{2}\rangle)=
\sum_{q^{'},\gamma_{1}^{'},\gamma_{2}^{'}}\delta^{*}(q^{'},\gamma_{1}^{'},\gamma_{2}^{'},
\sigma,q,\gamma_{1},\gamma_{2})
|q^{'}\rangle|\gamma_{1}^{'}\rangle|\gamma_{2}^{'}\rangle,
\end{equation}
where $\delta^{*}$ denotes the conjugate complex number $\delta$.
By linearity $U_{\sigma}$ and $U_{\sigma}^{'}$ can be extended to
$H_{Q}\otimes H_{\Gamma^{*}}\otimes H_{\Gamma^{*}}$.

{\bf Remark 1.} $U_{\sigma}^{'}$ is the adjoint operator of
$U_{\sigma}^{'}$. Indeed, for any
$(q_{i},\gamma_{i1},\gamma_{i2})\in Q\times \Gamma^{*}\times
\Gamma^{*}$, by means of Eqs. (1,2) we have
\begin{eqnarray}
&&\hskip -15mm\nonumber\left\langle
U_{\sigma}|q_{1}\rangle|\gamma_{11}\rangle|\gamma_{12}\rangle,
U_{\sigma}|q_{2}\rangle|\gamma_{21}\rangle|\gamma_{22}\rangle\right\rangle\\
&=&\nonumber
\sum_{q,\gamma_{1},\gamma_{2}}\delta(q_{1},\gamma_{11},\gamma_{12},\sigma,q,\gamma_{1},\gamma_{2})\times
\delta^{*}(q_{2},\gamma_{21},\gamma_{22},\sigma,q,\gamma_{1},\gamma_{2})\\
&=&\left\langle
|q_{1}\rangle|\gamma_{11}\rangle|\gamma_{12}\rangle,
U_{\sigma}^{'}U_{\sigma}|q_{2}\rangle|\gamma_{21}\rangle|\gamma_{22}\rangle\right\rangle.
\end{eqnarray}

{\bf Definition 2.} Let $M$ be a quasi-quantum two-stack machine
with input alphabet $\Sigma$. If $U_{\sigma}$ is unitary for any
$\sigma\in\Sigma\cup\{\#,\$\}$, then $M$ is called a quantum
two-stack machine.

Now we give the well-formedness conditions for justifying the
unitarity of $U_{\sigma}$ for any $\sigma\in\Sigma\cup\{\#,\$\} $,
that are described by the following theorem.

{\bf Theorem 1.} Let $M$ be a quasi-quantum two-stack machine with
input alphabet $\Sigma$. Then for any
$\sigma\in\Sigma\cup\{\#,\$\}$, linear operator $U_{\sigma}$ is
unitary if and only if $\delta$ satisfies the following
well-formedness conditions:

(I) For any $\sigma\in\Sigma\cup\{\#,\$\}$,
\begin{eqnarray}
&&\hskip
-15mm\nonumber\sum_{q^{'},\gamma_{1}^{'},\gamma_{2}^{'}}\delta(q_{1},\gamma_{11},\gamma_{12},\sigma,q^{'},\gamma_{1}^{'},\gamma_{2}^{'})\times
\delta^{*}(q_{2},\gamma_{21},\gamma_{22},\sigma,q^{'},\gamma_{1}^{'},\gamma_{2}^{'})\\
&=&\left\{\begin{array}{ll} 1,& {\rm if}\hskip 2mm
(q_{1},\gamma_{11},\gamma_{12})=(q_{2},\gamma_{21},\gamma_{22}),\\
0,& {\rm otherwise}.
\end{array}
\right.
\end{eqnarray}

(II) For any $\sigma\in\Sigma\cup\{\#,\$\}$,
\begin{eqnarray}
&&\hskip
-15mm\nonumber\sum_{q^{'},\gamma_{1}^{'},\gamma_{2}^{'}}\delta(q^{'},\gamma_{1}^{'},\gamma_{2}^{'},\sigma,q_{1},\gamma_{11},\gamma_{12})\times
\delta^{*}(q^{'},\gamma_{1}^{'},\gamma_{2}^{'},\sigma,q_{2},\gamma_{21},\gamma_{22})\\
&=&\left\{\begin{array}{ll} 1,& {\rm if}\hskip 2mm
(q_{1},\gamma_{11},\gamma_{12})=(q_{2},\gamma_{21},\gamma_{22}),\\
0,& {\rm otherwise}.
\end{array}
\right.
\end{eqnarray}

{\bf Proof.} Is is similar to Theorem 5 in [9], and the details
are referred to Appendix I. $\Box$

\section*{3. Quantum multi-counter machines}

QCAs were first considered by Kravtsev [6], and further developed
by Yamasaki et al. [10]. In this section, we introduce a different
definition of quantum $k$-counter machines, and then deal with
some simulations between QMCMs with different counters and with
different counts in each move.

{\bf Definition 3.} A quasi-quantum $k$-counter machine is defined
as $M=(Q,\Sigma,\delta,q_{0},q_{a},q_{r})$ where $Q$ is a set of
states with initial state $q_{0}\in Q$ and states $q_{a},q_{r}\in
Q $ representing accepting and rejecting states, respectively,
$\Sigma$ is an input alphabet, and transition function $\delta$ is
a mapping from

$Q\times {\bf N}^{k}\times (\Sigma\cup\{\#,\$\})\times Q\times {\bf N}^{k}$\\
to ${\bf C}$, where ${\bf N}$ denotes the set of all nonnegative
integer and $\#,\$$ represent two endmarkers that begins with $\#$
and ends with $\$$, and $\delta$ satisfies that
\begin{equation}
\delta(q,n_{1},n_{2},\ldots,n_{k},\sigma,q^{'},n_{1}^{'},n_{2}^{'},\ldots,n_{k}^{'})\not=0
\end{equation}
only if $|n_{i}-n_{i}^{'}|\leq 1$ for $i=1,2,\ldots,k$.
Furthermore, let $|q\rangle|n_{1}\rangle|n_{2}\rangle\ldots
|n_{k}\rangle$ represent a configuration of $M$, where $q\in Q$,
$n_{i}\in {\bf N}$ for $i=1,2,\ldots$, and let the set

$C_{M}=\{|q\rangle|n_{1}\rangle|n_{2}\rangle\ldots |n_{k}\rangle:
q\in Q, n_{i}\in {\bf N},i=1,2,\ldots,k\}$\\
be an orthonormal basis for the space $H_{C_{M}}=l_{2}(C_{M})$.
For any $\sigma\in\Sigma$, linear operator $V_{\sigma}$ on
$H_{C_{M}}$ is defined as follows:
\begin{equation}
V_{\sigma}|q\rangle|n_{1}\rangle|n_{2}\rangle\ldots
|n_{k}\rangle=\sum_{q^{'},n_{1}^{'},n_{2}^{'},\ldots,n_{k}^{'}}\delta(q,n_{1},n_{2},\ldots,n_{k},\sigma,q^{'},n_{1}^{'},n_{2}^{'},\ldots,n_{k}^{'})
|q^{'}\rangle|n_{1}^{'}\rangle|n_{2}^{'}\rangle\ldots
|n_{k}^{'}\rangle
\end{equation}
and $V_{\sigma}$ is extended to $H_{C_{M}}$ by linearity.

{\bf Definition 4.} We say that the quasi-quantum counter machine
$M=(Q,\Sigma,\delta,q_{0},q_{a},q_{r})$ defined above is a {\it
quantum $k$-counter machine}, if $V_{\sigma}$ is unitary for any
$\sigma\in (\Sigma\cup\{\#,\$\})$.

Also, we define linear operator $V_{\sigma}^{'}$ on $H_{C_{M}}$ as
follows:
\begin{equation}
V_{\sigma}^{'}|q\rangle|n_{1}\rangle|n_{2}\rangle\ldots
|n_{k}\rangle=\sum_{q^{'},n_{1}^{'},n_{2}^{'},\ldots,n_{k}^{'}}\delta^{*}(q^{'},n_{1}^{'},n_{2}^{'},\ldots,n_{k}^{'},\sigma,q,n_{1},n_{2},\ldots,n_{k})
|q^{'}\rangle|n_{1}^{'}\rangle|n_{2}^{'}\rangle\ldots
|n_{k}^{'}\rangle
\end{equation}

{\bf Remark 2.} Clearly $V_{\sigma}^{'}$ is an adjoint operator of
$V_{\sigma}$, which can be checked in terms of the process of
Remark 1, and the details are therefore omitted here.

Now we give the W-F conditions for characterizing the unitarity of
$V_{\sigma}$, that are described in the following theorem. For the
sake of simplicity, we deal with the case of $k=2$, and the other
cases are exactly similar.

{\bf Theorem 2.} Let $M$ be a quasi-quantum two-counter machine
with input alphabet $\Sigma$. Then for any
$\sigma\in\Sigma\cup\{\#,\$\}$, $V_{\sigma}$ defined as Eq. (7) is
unitary if and only if $\delta$ satisfies the following W-F
conditions:

(I) For any $\sigma\in\Sigma\cup\{\#,\$\}$,
\begin{eqnarray}
&&\hskip -15mm\nonumber
\sum_{p,n_{1}^{'},n_{2}^{'}}\delta(q_{1},n_{11},n_{12},\sigma,p,n_{1}^{'},n_{2}^{'})\times\delta^{*}(q_{2},n_{21},n_{22},\sigma,p,n_{1}^{'},n_{2}^{'})\\
&=&\left\{\begin{array}{ll} 1,& {\rm if} \hskip 2mm
(q_{1},n_{11},n_{12})=(q_{2},n_{21},n_{22}),\\
0,&{\rm otherwise}.
\end{array}
\right.
\end{eqnarray}

(II) For any $\sigma\in\Sigma\cup\{\#,\$\}$,
\begin{eqnarray}
&&\hskip -15mm\nonumber
\sum_{p,n_{1}^{'},n_{2}^{'}}\delta(p,n_{1}^{'},n_{2}^{'},\sigma,q_{1},n_{11},n_{12})\times\delta^{*}(p,n_{1}^{'},n_{2}^{'},\sigma,q_{2},n_{21},n_{22})\\
&=&\left\{\begin{array}{ll} 1,& {\rm if} \hskip 2mm
(q_{1},n_{11},n_{12})=(q_{2},n_{21},n_{22}),\\
0,&{\rm otherwise}.
\end{array}
\right.
\end{eqnarray}

{\bf Proof.} It is similar to Theorem 1 above, and the details are
presented in Appendix II. $\Box$

In order to simulate QTMs by QMSMs, we need some related lemmas
and definitions. In general, quantum counter machines are allowed
to count by $\pm 1$ and $0$ only. Here we would like to deal with
the quantum machines with count beyond such a bound, and show that
they are indeed equivalent.

{\bf Definition 5.} A quasi-quantum $k$-counter machine
$M=(Q,\Sigma,\delta,q_{0},q_{a},q_{r})$ is called to {\it count
with $\pm r$} for $r\geq 1$, if its $k$'s counters are allowed to
change with numbers $0,\pm 1$, or $\pm r$ at each step. In this
case, if $|n_{i}-n_{i}^{'}|\leq 1$ or $|n_{i}-n_{i}^{'}|=r$ for
$i=1,2,\ldots,k$, then

$\delta(q,n_{1},n_{2},\ldots,n_{k},\sigma,q^{'},n_{1}^{'},n_{2}^{'},\ldots,n_{k}^{'})\not=0$\\
may hold; otherwise it is 0. We say that the quasi-quantum
$k$-counter machine $M$ is quantum if for any $\sigma\in
\Sigma\cup\{\#,\$\}$, $V_{\sigma}$ is a unitary operator on
$l_{2}(C_{M})$, where
\[
C_{M}=\{|q\rangle|n_{1}\rangle|n_{2}\rangle\ldots
|n_{k}\rangle:q\in Q, n_{i}\in {\bf N}, i=1,2,\ldots,k\}.
\]

It is ready to obtain that Theorem 2 also holds for quantum
$k$-counter machines with count $\pm r$ for $r\geq 1$.

{\bf Theorem 3.} Let $M=(Q,\Sigma,\delta,q_{0},q_{a},q_{r})$ be a
quasi-quantum $k$-counter machine that is allowed to count with a
certain $\pm r$ for $r\geq 1$. Then for any $\sigma\in
\Sigma\cup\{\#,\$\}$, $V_{\sigma}$ defined as Eq. (7) is unitary
if and only if $\delta$ satisfies Eqs. (9,10).

{\bf Proof.} Similar to Theorem 2. $\Box$

{\bf Definition 6.} Let $M_{1}$ and $M_{2}$ be quantum
$k_{1}$-counter machine $M_{1}$ and quantum $k_{2}$-counter
machine $M_{2}$, respectively, and, $M_{1}$ and $M_{2}$ have the
same input alphabet $\Sigma$. For any
$\sigma\in\Sigma\cup\{\#,\$\}$, $V_{\sigma}^{(1)}$ and
$V_{\sigma}^{(2)}$ defined as Eq. (7) represent the evolution
operators in $M_{1}$ and $M_{2}$, respectively.  We say that
$M_{1}$ can simulate $M_{2}$, if for any string
$\sigma_{1}\sigma_{2}\ldots\sigma_{n}\in\Sigma^{*}$,
\begin{eqnarray}
&&\hskip -15mm\nonumber\sum_{i_{1},i_{2},\ldots,i_{k_{1}}\geq
0}\langle i_{k_{1}}|\ldots\langle i_{1}|\langle
q_{a}^{(1)}|V_{\$}^{(1)}
V_{\sigma_{n}}^{(1)}V_{\sigma_{n-1}}^{(1)}\ldots
V_{\sigma_{1}}^{(1)}V_{\#}^{(1)}|q_{0}^{(1)}\rangle|0\rangle\ldots|0\rangle\\
&=&\sum_{j_{1},j_{2},\ldots,j_{k_{1}}\geq 0}\langle
i_{k_{1}}|\ldots\langle i_{1}|\langle q_{a}^{(2)}|V_{\$}^{(2)}
V_{\sigma_{n}}^{(2)}V_{\sigma_{n-1}}^{(2)}\ldots
V_{\sigma_{1}}^{(2)}V_{\#}^{(2)}|q_{0}^{(2)}\rangle|0\rangle\ldots|0\rangle,
\end{eqnarray}
where $q_{0}^{(i)}$ and $q_{a}^{(i)}$ denote the initial and
accepting states of $M_{i}$, respectively, $i=1,2$.

For convenience, for any quantum $k$-counter machine
$M=(Q,\Sigma,\delta,q_{0},q_{a},q_{r})$, we define the accepting
probability $P_{accept}^{M}(\sigma_{1}\sigma_{2}\ldots\sigma_{n})$
for inputting $\sigma_{1}\sigma_{2}\ldots\sigma_{n}$ as:
\begin{eqnarray}
&&\hskip -15mm\nonumber P_{accept}^{M}(\sigma_{1}\sigma_{2}\ldots\sigma_{n})\\
&=&\sum_{i_{1},i_{2},\ldots,i_{k}\geq 0}\langle i_{k}|\ldots
\langle i_{1}|\langle q_{a}|V_{\$}^{M}
V_{\sigma_{n}}^{M}V_{\sigma_{n-1}}^{M}\ldots
V_{\sigma_{1}}^{M}V_{\#}^{M}|q_{0}\rangle|0\rangle\ldots
|0\rangle,
\end{eqnarray}
where $V_{\sigma}^{M}$ is unitary operator on $l_{2}(C_{M})$ for
any $\sigma\in \Sigma\cup\{\#,\$\}$.

The detailed proofs of the following two lemmas are given in
Appendix III.

{\bf Lemma 1.} For any quantum $k$-counter machine $M_{1}$ that is
allowed to count with $\pm r$ for $r\geq 1$, there exists quantum
$2k$-counter machine $M_{2}$ simulating $M_{1}$ with the same time
complexity, where $M_{2}$ is allowed to count with $0,\pm 1$, and
$\pm (r-1)$.

{\bf Lemma 2.} For any quantum $k$-counter machine $M_{1}$ that is
allowed to count with $0,\pm 1,\pm 2,\ldots,\pm r$, then there
exists a quantum $kr$-counter machine $M_{2}$ simulating $M_{1}$
with the same time complexity, where $M_{2}$ is allowed to count
with $0,\pm 1$ only.

\section*{4. Simulations of quantum Turing machines }

\subsection*{{\it 4.1. Simulations of quantum Turing machines in terms of quantum multi-counter
machines}}

To simulate QTMs in terms of QMCMs, we give the definition of QTMs
in terms of Bernstein and Vazirani [1], in which the read-write
head will move either to the right or to the left at each step.
Indeed, generalized QTMs can also be simulated by QMCMs, but the
discussion regarding unitarity is much more complicated. For the
sake of simplicity, we here consider the former QTMs.

{\bf Definition 7.} A QTM is defined by
$M=(\Sigma,Q,\delta,B,q_{0},q_{a},q_{r})$, where $\Sigma$ is a
finite input alphabet, $B$ is an identified blank symbol, $Q$ is a
finite set of states with an identified initial state $q_{0}$ and
final state $q_{a},q_{r}\not= q_{0}$, where $q_{a}$ and $q_{r}$
represent accepting and rejecting states, respectively, and the
quantum transition function $\delta$ is defined as
\[
\delta: Q\times\Sigma\times\Sigma\times Q\times\{L,R\}\rightarrow
{\bf C}.
\]
The QTM has a two-way infinite tape of cells indexed by ${\bf Z}$
and a single read-write tape head that moves along the tape. A
configuration of this machine is described by the form
$|q\rangle|\tau\rangle|i\rangle$, where $q$ denotes the current
state, $\tau\in \Sigma^{{\bf Z}}$ describes the tape symbols, and
$i\in {\bf Z}$ represents the current position of tape head.
Naturally, a configuration containing initial or final state is
called an initial or final configuration. Let $C_{M}$ denote the
set of all configurations in $M$, and therefore
$H_{C_{M}}=l_{2}(C_{M})$, that is a Hilbert space whose
orthonormal basis can be equivalently viewed as $C_{M}$. Then the
evolution operator $U_{M}$ on $l_{2}(C_{M})$ can be defined in
terms of $\delta$: for any configuration $|c\rangle\in C_{M}$,
\begin{equation}
U_{M}|c\rangle=\sum_{|c^{'}\rangle\in
C_{M}}a(c,c^{'})|c^{'}\rangle,
\end{equation}
where $a(c,c^{'})$ is the amplitude of configuration $|c\rangle$
evolving into $|c^{'}\rangle$ in terms of the transition function
$\delta$. $U_{M}$ is a unitary operator on $l_{2}(C_{M})$.

As in [1], we define that QTM halts with running time $T$ on input
$x$ if after the $T$'s step moves beginning with its initial
configuration, the superposition contains only final
configurations, and at any time less than $T$ the superposition
contains no final configuration. Therefore, we assume that the QTM
satisfies this requirement.

{\bf Definition 8.} For nonnegative integer $n,T$, let
$M_{1}=(Q_{1},\Sigma_{1},\delta_{1},B_{1},q_{10},q_{1a},q_{1r})$
be a quantum Turing machine with initial state $q_{10}$, and let
$M_{2}$ be a quantum $k$-counter machine with initial state
$q_{20}$ and the same input alphabet $\Sigma_{1}$ as $M_{1}$. We
say that $M_{2}$ $(n,T)$-simulates quantum Turing machine $M_{1}$
with {\it polynomial time $O(n,T)$ slowdown}, if there exist some
tape symbols added in $M_{2}$, say $B_{2},B_{3},\ldots, B_{m}$
such that for any input
$x=\sigma_{1}\sigma_{2}\ldots\sigma_{l}\in\Sigma_{2}^{*}$ $(l\leq
n)$, if the computation of $M_{1}$ ends with $t$ steps ($t\leq
T$), then there is nonnegative integers
$k_{l_{1}},k_{l_{2}},\ldots,k_{l_{m_{1}}}$ and
$k_{s_{1}},k_{s_{2}},\ldots,k_{s_{m_{2}}}$ that are related to $l$
and $t$, satisfying
$\sum_{i=1}^{m_{1}}k_{l_{i}}+\sum_{i=1}^{m_{2}}k_{s_{i}}\leq
O(n,T)$, and
\begin{equation}
P_{a}^{M_{1}}(x)=P_{a}^{M_{2}}(x),
\end{equation}
where
\begin{equation}
P_{a}^{M_{1}}(x)=\sum_{-T\leq i\leq T,\tau\in \Sigma^{[-T,T]_{\bf
Z} }}\left|\langle i|\langle\tau|\langle
q_{1a}|U_{M_{1}}^{t}|q_{10}\rangle |\tau_{0}\rangle
|0\rangle\right|^{2}
\end{equation}
where $\tau_{0}$ is defined as:
$\tau_{0}(j)=\left\{\begin{array}{ll} \sigma_{j+1},& {\rm
if} \hskip 2mm j\in [0,l-1]_{{\bf Z}},\\
B_{1},& j\in [-T,-1]_{{\bf Z}}\cup [l,T]_{{\bf Z}},
\end{array}
\right.$ and
\begin{eqnarray}
\hskip -25mm P_{a}^{M_{2}}(x)&=&\nonumber
\sum_{n_{1},n_{2},\ldots,n_{k}\geq 0}\left|\langle n_{k}|\ldots
\langle n_{1}|\langle
q_{2a}|V_{\$}V_{B_{s_{m_{2}}}}^{k_{s_{m_{2}}}}\ldots
V_{B_{s_{1}}}^{k_{s_{1}}}\right.\\
&&\hskip 10mm  \left. V_{\sigma_{l}}V_{\sigma_{l-1}}\ldots
V_{\sigma_{1}}V_{B_{l_{m_{1}}}}^{k_{l_{m_{1}}}}\ldots
V_{B_{l_{1}}}^{k_{l_{1}}}V_{\#}|q_{0}\rangle |0\rangle\ldots
|0\rangle\right|^{2}.
\end{eqnarray}

The main result of this subsection is as follows:

{\bf Theorem 4.} For any QTM
$M_{1}=(Q_{1},\Sigma_{1},\delta_{1},B_{1},q_{10},q_{1a},q_{1r})$
with initial state $q_{10}$ and accepting and rejecting states
$q_{1a},q_{1r}$, and for any nonnegative integer $n,t$ with $n\leq
t+1$, there exists a quantum (2$t$+2)-counter machine $M_{2}$ that
$(n,t)$-simulates $M_{1}$ with most slowdown $O(n+t)$.

{\bf Proof.} The details are referred to Appendix IV, but we
outline the basic idea as follows: We add three assistant input
symbols $B_{2},B_{3},B_{4}$ in $M_{2}$. For any input string
$\sigma_{1}\sigma_{2}\ldots\sigma_{k}\in\Sigma_{1}^{*}$ with
$k\leq n$, we take certain integer numbers
$l_{2}(k),l_{3}(k),l_{4}(k)$ that are related to $k$, and the
input string $\sigma_{1}\sigma_{2}\ldots\sigma_{k}$ is put on the
tape of $M_{2}$ in the form
$\#\sigma_{1}\sigma_{2}\ldots\sigma_{k}B_{2}^{l_{2}(k)}B_{3}^{l_{3}(k)}B_{4}^{l_{4}(k)}\$$.
After $M_{2}$ finishes the reading of $B_{3}^{l_{3}(k)}$, the
counters from $t+1$ to $t+k$ have numbers corresponding to
$\sigma_{1},\sigma_{2},\ldots,\sigma_{k}$, respectively, those
from $1$ to $t$ are numbers corresponding to blank $B_{1}$ in
$M_{1}$, and the last counter $2t+2$ is always used to simulate
the position of the read-write head of $M_{1}$. Assistant symbols
$B_{4}^{l_{4}(k)}$ are read one by one in the process of $M_{1}$
computing $\sigma_{1}\sigma_{2}\ldots\sigma_{k}$. When $M_{1}$
ends, $M_{2}$ will read $\$$ and stop. In order to preserve the
unitarity of $M_{2}$, some additional definitions for the
transition function $\delta_{2}$ of $M_{2}$ are necessary. $\Box$

\subsection*{{\it 4.2. Simulations of quantum Turing machines in terms of quantum
multi-stack machines}}

QTMs can be also $(n,t)$-simulated by quantum multi-stack machine,
since quantum $k$-counter machine can be simulated by quantum
multi-stack machine in terms of the following Theorem 5.

{\bf Definition 9.} We say that quantum $k$-stack machine
$M_{2}=(Q_{2},\Sigma_{2},\Gamma_{2},\delta_{2},Z_{0},q_{20},q_{2a},q_{2r})$
simulates quantum $k$-counter machine
$M_{1}=(Q_{1},\Sigma_{1},\delta_{1},q_{10},q_{1a},q_{1r})$ that
has the same input alphabet $\Sigma_{1}=\Sigma_{2}$ with the same
time complexity, if for any input string
$x=\sigma_{1}\sigma_{2}\ldots\sigma_{n}\in \Sigma_{1}^{*}$, we
have
\begin{equation}
P_{accept}^{M_{1}}(x)=P_{accept}^{M_{2}}(x)
\end{equation}
where
\begin{equation}
P_{accept}^{M_{1}}(x)=\nonumber\sum_{\gamma_{1},\gamma_{2},\ldots,\gamma_{k}}\left|\langle\gamma_{k}|\langle\gamma_{k-1}|\ldots
\langle \gamma_{1}|\langle q_{1a}|\right.  \left.
U_{\$}U_{\sigma_{n}}\ldots
U_{\sigma_{1}}U_{\#}|q_{10}\rangle|0\rangle\ldots
|0\rangle\right|^{2}.
\end{equation}

As well, the $(n,t)$-simulations of QTMs in terms of quantum
$k$-stack machine can be similarly defined as Definition 8, and we
leave out the details here.

{\bf Theorem 5.} For any given quantum $k$-counter machine
$M_{1}=(Q_{1},\Sigma_{1},\delta_{1},q_{10},q_{1a},q_{1r})$, there
exists quantum $k$-stack machine $M_{2}$ that simulates $M_{1}$
with the same time complexity.

{\bf Proof.} Let
$M_{2}=(Q_{2},\Sigma_{2},\delta_{2},q_{20},q_{2a},q_{2r})$ where
$Q_{2}=Q_{1}$, $\Sigma_{2}=\{Z_{0},X\}$, $q_{20}=q_{10}$,
$q_{2a}=q_{1a}$, $q_{2r}=q_{1r}$. Define mapping $m: {\bf
N}\rightarrow \Sigma_{2}^{*}$ as follows:

$m(n)=Z_{0}X^{|n|}$\\
where we denote $X^{0}=\epsilon$, that is, $Z_{0}X^{0}=Z_{0}$. We
define $\delta_{2}$ as follows:
\begin{eqnarray*}
&&\hskip -15mm\delta_{2}(q,Z_{0}X^{l_{1}},Z_{0}X^{l_{2}},\ldots,Z_{0}X^{l_{k}},\sigma,p,Z_{0}X^{l_{1}^{'}},Z_{0}X^{l_{2}^{'}},\ldots,Z_{0}X^{l_{k}^{'}})\\
&=&\delta_{1}(q,l_{1},l_{2},\ldots,l_{k},\sigma,p,l_{1}^{'},l_{2}^{'},\ldots,l_{k}^{'})
\end{eqnarray*}
for any $q,p\in Q_{1}$, $\sigma\in\Sigma_{1}\cup\{\#,\$\}$, and
$l_{i},l_{i}^{'}\in {\bf N}$. Then it is easy to check that
$\delta_{2}$ satisfies the W-F conditions Eqs. (4,5), and that
$M_{2}$ simulates $M_{1}$ step by step. This completes the proof.
$\Box$

{\bf Corollary 1.} For any $n,t\in {\bf N}$, and any QTM $M_{1}$,
there exists QMSM $M_{2}$ that simulates $M_{1}$ with slowdown
$O(n+t)$.

\section*{5. Concluding remarks}

The unitary evolution of quantum physics requires that quantum
computation should be necessarily time reversible (unitary). This
makes some simulations between quantum computing devices quite
complicated. Indeed, the unitarity is reflected by the W-F
conditions. The W-F conditions for these QMSMs and QMCMs defined
in this paper are more succinct than the W-F conditions for QCAs
introduced by Yamasaki {\it et al.} [10], but we note that the
transition functions in our quantum devices employ the whole
property of the symbols in the stacks or counters at each move. An
issue worthy of further consideration is to give also succinct W-F
conditions but yet more local transition functions for
characterizing the unitarity of these QMSMs and QMCMs defined in
this paper. Moreover, the relationships between QMCMs in the paper
and QCAs by Yamasaki {\it et al.} [10] still need to be further
clarified. Finally,  how to improve the $(n,t)$-simulations of
QTMs by QMCMs and QMSMs towards more general simulations and how
to decrease the number of counters of QMCMs for simulating QTMs
are also worth studying.

\section*{Acknowledgements}

This research is supported by the National Natural Science
Foundation (No. 90303024), and the Natural Science Foundation of
Guangdong Province (No. 020146, 031541) of China.

\newpage

\section*{Appendix I. The proof of Theorem 1}

 First we show that if $\delta$ satisfies the
well-formedness conditions (I) and (II) above, then for any
$\sigma\in\Sigma\cup\{\#,\$\}$, $U_{\sigma}$ is unitary. For any
$|q_{i},\rangle|\gamma_{i1}\rangle|\gamma_{i2}\rangle\in C_{M}$,
$i=1,2$, by condition (I) we have
\begin{eqnarray*}
&&\left\langle
U_{\sigma}|q_{1}\rangle|\gamma_{11}\rangle|\gamma_{12}\rangle,U_{\sigma}|q_{2}\rangle|\gamma_{21}\rangle|\gamma_{22}\rangle\right\rangle\\
&=&\left\langle\sum_{p_{1},\gamma_{11}^{'},\gamma_{12}^{'}}\delta(q_{1},\gamma_{11},\gamma_{12},\sigma,p_{1},\gamma_{11}^{'},\gamma_{12}^{'})
|p_{1}\rangle|\gamma_{11}^{'}\rangle|\gamma_{12}^{'}\rangle,\right.\\
&&\left.\sum_{p_{2},\gamma_{21}^{'},\gamma_{22}^{'}}\delta(q_{2},\gamma_{21},\gamma_{22},\sigma,p_{2},\gamma_{21}^{'},\gamma_{22}^{'})
|p_{2}\rangle|\gamma_{21}^{'}\rangle|\gamma_{22}^{'}\rangle\right\rangle\\
&=&\sum_{q,\gamma_{1},\gamma_{2}}\delta(q_{1},\gamma_{11},\gamma_{12},\sigma,q,\gamma_{1},\gamma_{2})\times
\delta^{*}(q_{2},\gamma_{21},\gamma_{22},\sigma,q,\gamma_{1},\gamma_{2})\\
&=&\left\{\begin{array}{ll} 1,& {\rm if}\hskip 2mm
(q_{1},\gamma_{11},\gamma_{12})=(q_{2},\gamma_{21},\gamma_{22}),\\
0,& {\rm otherwise}.
\end{array}
\right.
\end{eqnarray*}
Furthermore, for any $\sigma\in\Sigma\cup\{\#,\$\}$, linear
operator $U_{\sigma}^{'}$ on $H_{Q}\otimes H_{\Gamma^{*}}\otimes
H_{\Gamma^{*}}$ is defined as Eq. (2). Then for any
$|q\rangle|\gamma_{1}\rangle|\gamma_{2}\rangle\in C_{M}$, with
condition (II) (Eq. (4)) we have
\begin{eqnarray*}
&&U_{\sigma}U_{\sigma}^{'}|q\rangle|\gamma_{1}\rangle|\gamma_{2}\rangle\\
&=&\sum_{q^{'},\gamma_{1}^{'},\gamma_{2}^{'}}\delta^{*}(q^{'},\gamma_{1}^{'},\gamma_{2}^{'},\sigma,q,\gamma_{1},\gamma_{2})
U_{\gamma}|q^{'}\rangle|\gamma_{1}^{'}\rangle|\gamma_{2}^{'}\rangle\\
&=&\sum_{q^{'},\gamma_{1}^{'},\gamma_{2}^{'}}\delta^{*}(q^{'},\gamma_{1}^{'},\gamma_{2}^{'},\sigma,q,\gamma_{1},\gamma_{2})\times
\sum_{q^{''},\gamma_{1}^{''},\gamma_{2}^{''}}\delta(q^{'},\gamma_{1}^{'},\gamma_{2}^{'},\sigma,q^{''},\gamma_{1}^{''},\gamma_{2}^{''})
|q^{''}\rangle|\gamma_{1}^{''}\rangle|\gamma_{2}^{''}\rangle\\
&=&|q\rangle|\gamma_{1}\rangle|\gamma_{2}\rangle,
\end{eqnarray*}
and similarly,
\[
U_{\sigma}^{'}U_{\sigma}|q\rangle|\gamma_{1}\rangle|\gamma_{2}\rangle=|q\rangle|\gamma_{1}\rangle|\gamma_{2}\rangle.
\]
So, we have verified that
$U_{\sigma}U_{\sigma}^{'}=U_{\sigma}^{'}U_{\sigma}=I$, and
therefore, $U_{\sigma}^{'}=U_{\sigma}^{-1}$. Therefore
$U_{\sigma}^{'}$ is also surjective and $U_{\sigma}$ has been
shown to be unitary. On the other hand, if $U_{\sigma}$ is
unitary, then
$U_{\sigma}U_{\sigma}^{'}=U_{\sigma}^{'}U_{\sigma}=I$ for any
$\sigma\in\Sigma\cup\{\#,\$\}$, since $U_{\sigma}^{'}$ is the
adjoint operator of $U_{\sigma}$. We need to demonstrate that the
well-formedness conditions (I) and (II) hold. Indeed, the
unitarity of $U_{\sigma}$ implies that for any
$|q_{i},\rangle|\gamma_{i1}\rangle|\gamma_{i2}\rangle\in C_{M}$,
$i=1,2$,
\begin{eqnarray*}
&&\sum_{q^{'},\gamma_{1}^{'},\gamma_{2}^{'}}\delta(q_{1},\gamma_{11},\gamma_{12},\sigma,q^{'},\gamma_{1}^{'},\gamma_{2}^{'})\times
\delta^{*}(q_{2},\gamma_{21},\gamma_{22},\sigma,q^{'},\gamma_{1}^{'},\gamma_{2}^{'})\\
&=&\left\langle
U_{\sigma}|q_{1}\rangle|\gamma_{11}\rangle|\gamma_{12}\rangle,U_{\sigma}|q_{2}\rangle|\gamma_{21}\rangle|\gamma_{122}\rangle\right\rangle\\
&=&\left\langle
|q_{1}\rangle|\gamma_{11}\rangle|\gamma_{12}\rangle,|q_{2}\rangle|\gamma_{21}\rangle|\gamma_{122}\rangle\right\rangle\\
&=&\left\{\begin{array}{ll} 1,& {\rm if}\hskip 2mm
(q_{1},\gamma_{11},\gamma_{12})=(q_{2},\gamma_{21},\gamma_{22}),\\
0,& {\rm otherwise}.
\end{array}\right.
\end{eqnarray*}
Furthermore, since
$U_{\sigma}U_{\sigma}^{'}=U_{\sigma}^{'}U_{\sigma}=I$,
$U_{\sigma}^{'}$ is also unitary. Therefore, the unitarity of
$U_{\sigma}^{'}$ implies that the condition (II) holds true, as
well. $\Box$

\section*{Appendix II. The proof of Theorem 2}

If $\delta$ satisfies the conditions (I) and (II) given by Eqs.
(9,10), then for any configurations
$|q\rangle|n_{1}\rangle|n_{2}\rangle$, in light of condition (I)
we have
\begin{eqnarray*}
V_{\sigma}^{'}V_{\sigma}|q\rangle|n_{1}\rangle|n_{2}\rangle&=&
V_{\sigma}^{'}\left(\sum_{q^{'},n_{1}^{'},n_{2}^{'}}\delta(q,n_{1},n_{2},\sigma,q^{'},n_{1}^{'},n_{2}^{'})|q^{'}\rangle|n_{1}^{'}\rangle|n_{2}^{'}\rangle\right)\\
&=&\sum_{q^{'},n_{1}^{'},n_{2}^{'}}\delta(q,n_{1},n_{2},\sigma,q^{'},n_{1}^{'},n_{2}^{'})\sum_{p,m_{1},m_{2}}\delta^{*}(p,m_{1},m_{2},\sigma,q^{'},n_{1}^{'},n_{2}^{'})
|p\rangle|m_{1}\rangle|m_{2}\rangle\\
&=&\sum_{q^{'},n_{1}^{'},n_{2}^{'}}\delta(q,n_{1},n_{2},\sigma,q^{'},n_{1}^{'},n_{2}^{'})\delta^{*}(q,n_{1},n_{2},\sigma,q^{'},n_{1}^{'},n_{2}^{'})
|q\rangle|n_{1}\rangle|n_{2}\rangle\\
&=&|q\rangle|n_{1}\rangle|n_{2}\rangle,
\end{eqnarray*}
and with condition (II) we have
\begin{eqnarray*}
V_{\sigma}V_{\sigma}^{'}|q\rangle|n_{1}\rangle|n_{2}\rangle&=&
V_{\sigma}\left(\sum_{q^{'},n_{1}^{'},n_{2}^{'}}\delta^{*}(q^{'},n_{1}^{'},n_{2}^{'},\sigma,q,n_{1},n_{2})|q^{'}\rangle|n_{1}^{'}\rangle|n_{2}^{'}\rangle\right)\\
&=&\sum_{q^{'},n_{1}^{'},n_{2}^{'}}\delta^{*}(q^{'},n_{1}^{'},n_{2}^{'},\sigma,q,n_{1},n_{2})\sum_{p,m_{1},m_{2}}
\delta(q^{'},n_{1}^{'},n_{2}^{'},\sigma,p,m_{1},m_{2})
|p\rangle|m_{1}\rangle|m_{2}\rangle\\
&=&\sum_{q^{'},n_{1}^{'},n_{2}^{'}}\delta^{*}(q^{'},n_{1}^{'},n_{2}^{'},\sigma,q,n_{1},n_{2})\delta(q^{'},n_{1}^{'},n_{2}^{'},\sigma,q,n_{1},n_{2},)
|q\rangle|n_{1}\rangle|n_{2}\rangle\\
&=&|q\rangle|n_{1}\rangle|n_{2}\rangle.
\end{eqnarray*}
Therefore, $V_{\sigma}^{'}V_{\sigma}=V_{\sigma}V_{\sigma}^{'}=I$
for any $\sigma\in\Sigma\cup\{\#,\$\}$, and $V_{\sigma}$ is
therefore unitary.

On the other hand, if $V_{\sigma}$ is unitary, then for any
$(q_{i},n_{i1},n_{i2})\in Q\times {\bf N}\times {\bf N}$, $i=1,2$,
we have
\begin{eqnarray*}
&&\left\langle
|q_{1}\rangle|n_{11}\rangle|n_{12}\rangle,|q_{2}\rangle|n_{21}\rangle|n_{22}\rangle\right\rangle\\
&=&\left\langle
V_{\sigma}|q_{1}\rangle|n_{11}\rangle|n_{12}\rangle,V_{\sigma}|q_{2}\rangle|n_{21}\rangle|n_{22}\rangle\right\rangle\\
&=&\sum_{p,n_{1}^{'},n_{2}^{'}}\delta(q_{1},n_{11},n_{12},\sigma,p,n_{1}^{'},n_{2}^{'})\times
\delta^{*}(q_{2},n_{21},n_{22},\sigma,p,n_{1}^{'},n_{2}^{'})\\
&=&\left\{\begin{array}{ll} 1,& {\rm if} \hskip 2mm
(q_{1},n_{11},n_{12})=(q_{2},n_{21},n_{22}),\\
0,& {\rm otherwise}.
\end{array}
\right.
\end{eqnarray*}
As well, the unitarity of $V_{\sigma}^{'}$ implies that condition
(II) holds. $\Box$

\section*{Appendix III. The proofs of Lemmas 1 and 2}

\subsection*{{\it The proof of Lemma 1:}}

We here consider the case of $k=1$ without loss of generality,
since it is similar to show the general situation. Suppose
$M_{1}=(Q_{1},\Sigma_{1},\delta_{1},q_{10},q_{1a},q_{1r})$. We
define a desired quantum two-counter machine $M_{2}$ simulating
$M_{1}$. A basic idea is that one of counters in $M_{2}$ simulates
the changes of $d_{1}\in\{-1,0,1\}$ in $M_{1}$ and the other
counter of $M_{2}$ simulates the changes of $d_{2}\in\{-r,0,r\}$
in $M_{1}$. More formally, we define
$M_{2}=(Q_{2},\Sigma_{2},\delta_{2},q_{20},q_{2a},q_{2r})$ as
follows: $Q_{2}=Q_{1}$, $\Sigma_{2}=\Sigma_{1}$, $q_{10}=q_{20}$,
$q_{2a}=q_{1a}$, $q_{2r}=q_{1r}$, and

$\delta_{2}:Q_{2}\times [0,r-1]_{{\bf Z}}\times {\bf N}\times\Sigma\times Q_{2}\times [0,r-1]_{{\bf Z}}\times {\bf N}\rightarrow {\bf C}$\\
where $[0,r-1]_{{\bf Z}}$ denotes the set $\{0,1,2,\ldots,r-1\}$.
Specifically, $\delta_{2}$ is defined in the following way: If
\begin{equation}
\delta_{1}(q,n_{1}+k_{1}r,\sigma,p,n_{2}+k_{2}r)=c
\end{equation}
where $0\leq n_{1},n_{2}<r$, $k_{1},k_{2}\geq 0$, and $c\in {\bf
C}$ denotes its amplitude, then
\begin{equation}
\delta_{2}(q,n_{1},k_{1},\sigma,p,n_{2},k_{2})=c.
\end{equation}
For example, if $\delta_{1}(q,kr+r-1,\sigma,p,(k+1)r)=c$, then
$\delta_{2}(q,r-1,k,\sigma,p,0,k+1)=c$; if
$\delta_{1}(q,kr,\sigma,p,kr-1)=c$, then
$\delta_{2}(q,0,k,\sigma,p,r-1,k-1)=c$.

In terms of the definition of $\delta_{2}$ as Eqs. (19,20), we
have also defined a linear operator $V_{\sigma}^{M_{2}}$ by Eq.
(7) on space $l_{2}(C_{M_{2}})=H_{Q_{2}}\otimes H_{[0,r-1]_{{\bf
Z}} }\otimes H_{{\bf N}}$, where $H_{X}$, as above, is identified
with a Hilbert space whose orthonormal basis is the set $X$, and
$C_{M_{2}}$ is the set of configurations of $M_{2}$, as follows:

$C_{M_{2}}=\{|q\rangle|i\rangle|j\rangle: q\in Q_{2},0\leq i\leq
r-1, j\geq
0\}$.\\
By means of Eqs. (19,20), we will show that $\delta_{1}$
satisfying Eqs. (9,10) implies that $\delta_{2}$ also satisfies
Eqs. (9,10), and therefore, by Theorem 3 $V_{\sigma}^{M_{2}}$ is a
unitary operator on $l_{2}(C_{M_{2}})$. Furthermore, for
configuration $|q\rangle|i\rangle|j\rangle$ with $i\geq r$, we may
define

$V_{\sigma}^{M_{2}}|q\rangle|i\rangle|j\rangle=|q\rangle|i\rangle|j\rangle$,\\
then $V_{\sigma}^{M_{2}}$ is exactly extended to be a unitary
operator on $H_{Q_{2}}\otimes H_{{\bf N}}\otimes H_{{\bf N}}$.

More precisely, we show that $V_{\sigma}^{M_{2}}$ is a unitary
operator on $l_{2}(C_{M_{2}})=H_{Q_{2}}\otimes H_{[0,r-1]_{{\bf
Z}} }\otimes H_{{\bf N}}$. For any $(q_{i},n_{i1},n_{i2})\in
Q_{2}\times [0,r-1]_{{\bf Z}}\times {\bf N}$, $i=1,2$,
\begin{eqnarray}
&&\nonumber\sum_{p,n_{1},n_{2}}\delta_{2}(q_{1},n_{11},n_{12},\sigma,p,n_{1},n_{2})\times\delta_{2}^{*}(q_{2},n_{21},n_{22},\sigma,p,n_{1},n_{2})\\
&=&\nonumber\sum_{p,n_{1},n_{2}}\delta_{1}(q_{1},n_{11}+n_{12}r,\sigma,p,n_{1}+n_{2}r)\times\delta_{1}^{*}(q_{2},n_{21}+n_{22}r,\sigma,p,n_{1}+n_{2}r)\\
&=&\left\{\begin{array}{ll} 1,& {\rm if}\hskip 2mm
(q_{1},n_{11},n_{12})=(q_{2},n_{21},n_{22}),\\
0,&{\rm otherwise}.
\end{array}
\right.
\end{eqnarray}
Similarly, we have that for any $(p_{i},n_{i1},n_{i2})\in Q\times
[0,r-1]_{{\bf Z}}\times {\bf N}$, $i=1,2$,
\begin{eqnarray}
&&\nonumber\sum_{q,n_{1},n_{2}}\delta_{2}(q,n_{1},n_{2},\sigma, p_{1},n_{11},n_{12})\times\delta_{2}^{*}(q,n_{1},n_{2},\sigma,p,n_{21},n_{22})\\
&=&\nonumber\sum_{q,n_{1},n_{2}}\delta_{1}(q,n_{1}+n_{2}r,\sigma,p_{1},n_{11}+n_{12}r)\times\delta_{1}^{*}(q,n_{1}+n_{2}r,\sigma,p_{2},n_{21}+n_{22}r)\\
&=&\left\{\begin{array}{ll} 1,& {\rm if}\hskip 2mm
(p_{1},n_{11},n_{12})=(p_{2},n_{21},n_{22}),\\
0,& {\rm otherwise}.
\end{array}
\right.
\end{eqnarray}
Therefore, by Theorem 3 $V_{\sigma}^{M_{2}}$ is unitary for any
$\sigma\in\Sigma_{2}$. The remainder is to show that
$P_{accept}^{M_{1}}(x)=P_{accept}^{M_{2}}(x)$ for any
$x\in\Sigma_{1}^{*}$, which follows from Eqs. (19,20) and the
definition of $P_{accept}^{M}(x)$ described by Eq. (12). $\Box$

\subsection*{{\it The proof of Lemma 2:}}

We prove the case of $k=1$ without loss of generality. Let
$M_{1}=(Q_{1},\Sigma_{1},\delta_{1},q_{10},q_{1a},q_{1r})$ be a
quantum 1-counter machine. Then quantum $r$-counter machine
$M_{2}=(Q_{2},\Sigma_{2},\delta_{2},q_{20},q_{2a},q_{2r})$ is
defined as: $Q_{2}=Q_{1}$, $\Sigma_{2}=\Sigma_{1}$,
$q_{20}=q_{10}$, $q_{2a}=q_{1a}$, $q_{2r}=q_{1r}$,  and
$\delta_{2}$ is defined as follows: If
\begin{equation}
\delta_{1}(q,k_{1}r+i_{1},\sigma,p,k_{2}r+i_{2})=c,
\end{equation}
where $0\leq i_{1},i_{2}\leq r-1$ and $k_{1},k_{2}\geq 0$,  then
we define
\begin{equation}
\delta_{2}(q,(\underbrace{0,0,\ldots,0,1}_{i_{1}},0,\ldots,0,k_{1}),\sigma,p,(\underbrace{0,0,\ldots,0,1}_{i_{2}},0,\ldots,0,k_{2}))=c,
\end{equation}
where $\delta_{2}$ satisfies that if $k_{2}-k_{1}=1$ then
$i_{2}\leq i_{1}$; if $k_{2}-k_{1}=-1$, then $i_{2}\geq i_{1}$;
$\underbrace{0,0,\ldots,1}_{i_{1}}$ denotes that the $i_{1}$th
number is $1$, the $j$th number is 0 for $1\leq j\leq i_{1}-1$,
and, $(\underbrace{0,0,\ldots,0,1}_{i_{1}},0,\ldots,0,k_{1})$
represents that the $r$th number is $k_{1}$. Therefore the
definition by Eq. (24) describes equally the amplitude that in
$M_{1}$ the current number in the counter is $k_{1}r+i_{1}$ and
reading $\sigma$ leads to the number becoming $k_{2}r+i_{2}$. As
well, it is easy to see that if in $M_{1}$ the count changes with
numbers $0,\pm 1,\pm 2,\ldots,\pm r$, then in $M_{2}$ the count
changes with $0,\pm 1$. Furthermore, we show that the unitarity of
$M_{1}$ leads to the evolution operator $V_{\sigma}^{M_{2}}$
defined by Eq. (7) in $M_{2}$ being unitary for any
$\sigma\in\Sigma_{2}\cup \{\#,\$\}$. Indeed, $\delta_{2}$ is a
mapping on set

$\left\{(q,n_{1},n_{2},\ldots,n_{r}): q\in Q, n_{i}\in {\bf N},
i=1,2,\ldots,r, \sum_{i=1}^{r-1}n_{i}\leq 1\right\}$\\
and therefore, $V_{\sigma}$ is a linear operator on
$l_{2}(C_{M_{2}})$ where

$C_{M_{2}}=\left\{|q\rangle|n_{1}\rangle|n_{2}\rangle\ldots
|n_{r}\rangle: q\in Q, n_{i}\in {\bf N},
i=1,2,\ldots,r,\sum_{j=1}^{r-1}n_{j}\leq 1\right\}$,\\
that is, $l_{2}(C_{M_{2}})=H_{Q}\otimes (H_{\{0,1\}})^{\otimes
(r-1)}\otimes H_{{\bf N}}$.

For any $|q_{i}\rangle |n_{i1},n_{i2},\ldots,n_{ir}\rangle\in
C_{M_{2}}$, $i=1,2$,  then
\begin{eqnarray*}
&&\sum_{p,n_{1}^{'},n_{2}^{'},\ldots,n_{r}^{'},\sum_{i=1}^{r-1}n_{i}^{'}\leq
1}\delta_{2}(q_{1},n_{11},n_{12},\ldots,n_{1r},\sigma,p,n_{1}^{'},n_{2}^{'},\ldots,n_{r}^{'})\\
&&\hskip 30mm\times
\delta^{*}(q_{2},n_{21},n_{22},\ldots,n_{2r},\sigma,p,n_{1}^{'},n_{2}^{'},\ldots,n_{r}^{'})\\
&=&\sum_{p,n_{1}^{'},n_{2}^{'},\ldots,n_{r}^{'},\sum_{i=1}^{r-1}n_{i}^{'}\leq
1}\delta_{1}(q_{1},\sum_{i=1}^{r}n_{1i}i,\sigma,p,\sum_{i=1}^{r}n_{i}^{'}i)\times
\delta_{1}^{*}(q_{2},\sum_{i=1}^{r}n_{2i}i,\sigma,p,\sum_{i=1}^{r}n_{i}^{'}i)\\
&=&\sum_{p,n}\delta_{1}(q_{1},\sum_{i=1}^{r}n_{1i}i,\sigma,p,n)\times
\delta_{1}^{*}(q_{2},\sum_{j=1}^{r}n_{2j}j,\sigma,p,n)\\
&=&\left\{\begin{array}{ll} 1,& {\rm if}\hskip 2mm
(q_{1},\sum_{i=1}^{r}n_{1i}i)=(q_{2},\sum_{i=1}^{r}n_{2i}i)\hskip
2mm {\rm with}\hskip 2mm \sum_{j=1}^{r-1}n_{ij}\leq
1 \hskip 2mm {\rm for} \hskip 2mm i=1,2,\\
0,& {\rm otherwise},
\end{array}
\right.\\
&=&\left\{\begin{array}{ll} 1,& {\rm if}\hskip 2mm
(q_{1},n_{11},n_{12},\ldots,n_{1r})=(q_{2},n_{21},n_{22},\ldots,n_{2r}),\\
0,& {\rm otherwise}.
\end{array}
\right.
\end{eqnarray*}
As well, we have
\begin{eqnarray*}
&&\sum_{p,n_{1}^{'},n_{2}^{'},\ldots,n_{r}^{'},\sum_{i=1}^{r-1}n_{i}^{'}\leq
1}\delta_{2}(p,n_{1}^{'},n_{2}^{'},\ldots,n_{r}^{'},\sigma,q_{1},n_{11},n_{12},\ldots,n_{1r})\\
&&\hskip 30mm \times
\delta^{*}(p,n_{1}^{'},n_{2}^{'},\ldots,n_{r}^{'},\sigma,q_{2},n_{21},n_{22},\ldots,n_{2r})\\
&=&\left\{\begin{array}{ll} 1,& {\rm if}\hskip 2mm
(q_{1},n_{11},n_{12},\ldots,n_{1r})=(q_{2},n_{21},n_{22},\ldots,n_{2r}),\\
0,& {\rm otherwise}.
\end{array}
\right.
\end{eqnarray*}
Therefore, $V_{\sigma}^{M_{2}}$ is a unitary operator on
$l_{2}(C_{M_{2}})$. Furthermore, we denote $C_{M_{2}}^{'}$ by

$C_{M_{2}}^{'}=\{|q\rangle|n_{1}\rangle|n_{2}\rangle\ldots
|n_{r}\rangle: q\in Q, n_{i}\in {\bf N}, i=1,2,\ldots,r\}$. \\
Then we can extend $V_{\sigma}^{M_{2}}$ to be a unitary operator
on $l_{2}(C_{M_{2}}^{'})$ by defining as follows: for any
$|q\rangle|n_{1}\rangle|n_{2}\rangle\ldots |n_{r}\rangle\in
C_{M_{2}}^{'}\backslash C_{M_{2}}$,
\begin{eqnarray*}
&&\delta_{2}(q,n_{1},n_{2},\ldots,n_{r},\sigma,q^{'},n_{1}^{'},n_{2}^{'},\ldots,n_{r}^{'})\\
&=&\left\{\begin{array}{ll} 1,& {\rm if}\hskip 2mm
(q,n_{1},n_{2},\ldots,n_{r})=q^{'},n_{1}^{'},n_{2}^{'},\ldots,n_{r}^{'}),\\
0,& {\rm otherwise},
\end{array}
\right.
\end{eqnarray*}
and therefore,

$V_{\sigma}^{M_{2}}|q\rangle|n_{1}\rangle|n_{2}\rangle\ldots
|n_{r}\rangle=|q\rangle|n_{1}\rangle|n_{2}\rangle\ldots
|n_{r}\rangle$\\
for any $|q\rangle|n_{1}\rangle|n_{2}\rangle\ldots
|n_{r}\rangle\in C_{M_{2}}^{'}\backslash C_{M_{2}}$.

Finally, we show that for any $x\in\Sigma_{1}^{*}$,
\begin{equation}
P_{accept}^{M_{1}}(x)=P_{accept}^{M_{2}}(x).
\end{equation}
For any $\sigma\in\Sigma_{1}\cup\{\#,\$\}$, and $k\geq 0$, $0\leq
i\leq r-1$,
\begin{eqnarray}
&&\nonumber V_{\sigma}^{M_{1}}|q\rangle|kr+i\rangle\\
 &=&\nonumber
\sum_{p,|k^{'}r+i^{'}-kr-i|\leq
r}\delta_{1}(q,kr+i,\sigma,p,k^{'}r+i^{'})|p\rangle|k^{'}r+i^{'}\rangle\\
&=&\sum_{p,(k^{'},i^{'})\in
S(k,i)}\delta_{2}(q,\underbrace{0,\ldots,0,1}_{i},0,\ldots,0,k,\sigma,\underbrace{0,\ldots,0,1}_{i^{'}},0,\ldots,0,k^{'})
|p\rangle|k^{'}r+i^{'}\rangle,
\end{eqnarray}
where $S(k,i)=\{(k^{'},i^{'}):k^{'}=k,0\leq i^{'}\leq r-1, {\rm
or} \hskip 2mm 0\leq k^{'}=k-1,r-1\geq i^{'}\geq i, {\rm or}
\hskip 2mm k^{'}=k+1, r-1\leq i^{'}\leq i\}$. For any
$|q\rangle|n_{1}\rangle\ldots |n_{r}\rangle\in C_{M_{2}}$,
\begin{eqnarray}
&&\nonumber V_{\sigma}^{M_{2}}|q\rangle|n_{1}\rangle\ldots
|n_{r}\rangle\\
&=&\nonumber\sum_{q^{'},\sum_{i=1}^{r-1}n_{i}^{'}\leq
1}\delta_{2}(q,n_{1},n_{2},\ldots,n_{r},\sigma,n_{1}^{'},\ldots,n_{r}^{'})
|q^{'}\rangle |n_{1}^{'}\rangle\ldots |n_{r}^{'}\rangle\\
&=&\sum_{q^{'},\sum_{i=1}^{r-1}n_{i}^{'}\leq
1}\delta_{1}(q,\sum_{i=1}^{r}n_{i}i,\sigma,q^{'},\sum_{j=1}^{r}n_{j}^{'}j)|q^{'}\rangle
|n_{1}^{'}\rangle\ldots |n_{r}^{'}\rangle.
\end{eqnarray}
Therefore, Eq. (25) exactly follows from the definitions of
$V_{\sigma}^{M}$ and $P_{accept}^{M}$, and Eqs. (26,27) above.
$\Box$

\section*{Appendix IV. The proof of Theorem 4}

Let $|\Sigma_{1}|=r-2$. Then we define a quantum $2(t+1)$-counter
machine $M_{2}=(Q_{2},\Sigma_{2},\delta_{2},q_{0},q_{2a},q_{2r})$
that is allowed to count at each move with numbers $0,\pm
1,\ldots,\pm r$ simulating $M_{1}$, where $Q_{2}=Q_{1}\cup
\{q_{0},q_{0}^{'}\}$ with $q_{0},q_{0}^{'}\not\in Q_{1}$,
$q_{2a}=q_{1a}$, $q_{2r}=q_{1r}$, $\Sigma_{2}=\Sigma_{1}$, and
$\delta_{2}$ is defined as follows: Suppose that
$\Sigma_{1}=\{\sigma_{1},\sigma_{2},\ldots,\sigma_{r-2}\}$ and
bijective mapping $e: \Sigma_{1}\cup \{B_{1}\}\rightarrow
\{1,2,\ldots,r-1\}$.

(i) For any $\sigma\in\Sigma_{1}$, $r-1\geq n_{i}\geq 1$, $1\leq
i\leq 2t+1$, we define

1)
\begin{equation}
\delta_{2}(q_{0},0,\ldots,0,\#,q_{0},0,\ldots,0,1)=1;
\end{equation}
\begin{equation}
\delta_{2}(q_{0},0,\ldots,1,\#,q_{0},0,\ldots,0,0)=1;
\end{equation}

2)
\begin{equation}
\delta_{2}(q_{0},n_{1},n_{2},\ldots,n_{i-1},0,\ldots,0,i,\sigma,q_{0},n_{1},n_{2},\ldots,n_{i-1},e(\sigma),0,\ldots,0,i+1)=1;
\end{equation} for instance,
$\delta_{2}(q_{0},0,\ldots,0,1,\sigma,q_{0},k(\sigma),0,\ldots,0,2)=1$;
and when $i=2t+2$, we define
\begin{equation}
\delta_{2}(q_{0},n_{1},n_{2},\ldots,n_{2t+1},2t+2,\sigma,q_{0},0,\ldots,0,2t+2)=1,
\end{equation}
and for any $2\leq j\leq 2t+2$
\begin{equation}
\delta_{2}(q_{0},0,\ldots,0,j,\sigma,q_{0},0,\ldots,0,j-1)=1;
\end{equation}

 3) for any $1\leq l\leq n$, and  $1\leq i\leq l+1$, then
\begin{equation}
\delta_{2}(q_{0},n_{1},n_{2},\ldots,n_{l},0,\ldots,0,i,B_{2},q_{0},n_{1},n_{2},\ldots,n_{l},0,\ldots,0,i-1)=1;
\end{equation}
\begin{equation}
\delta_{2}(q_{0},n_{1},n_{2},\ldots,n_{l},0,\ldots,0,0,B_{2},q_{0},\underbrace{0,\ldots,0,n_{1}}_{t+1},n_{2},\ldots,n_{l},0,\ldots,0)=1;
\end{equation}
for $0\leq s\leq l$,
\begin{equation}
\delta_{2}(q_{0},\underbrace{0,\ldots,0,n_{1}}_{t+1},n_{2},\ldots,n_{l},0,\ldots,0,s,B_{2},
q_{0},\underbrace{0,\ldots,0,n_{1}}_{t+1},n_{2},\ldots,n_{l},0,\ldots,0,s+1)=1,
\end{equation}
and
\begin{equation}
\delta_{2}(q_{0},\underbrace{0,\ldots,0,n_{1}}_{t+1},n_{2},\ldots,n_{l},0,\ldots,0,l+1,B_{2},
q_{0},n_{1},n_{2},\ldots,n_{l},0,\ldots,0,l+1)=1.
\end{equation}

4) for any $1\leq l\leq n$, and  $0\leq i\leq t$, then
\begin{equation}
\delta_{2}(q_{0},\underbrace{0,\ldots,0,n_{1}}_{t+1},n_{2},\ldots,n_{l},0,\ldots,i,B_{3},
q_{0},\underbrace{0,\ldots,0,n_{1}}_{t+1},n_{2},\ldots,n_{l},0,\ldots,i+1)=1;
\end{equation}
\begin{eqnarray}
&&\nonumber\delta_{2}(q_{0},\underbrace{0,\ldots,0,n_{1}}_{t+1},n_{2},\ldots,n_{l},0,\ldots,t+1,B_{3},\\
&&\hskip 10mm
q_{0}^{'},\underbrace{e(B_{1}),\ldots,e(B_{1}),n_{1}}_{t+1},
n_{2},\ldots,n_{l},e(B_{1}),\ldots,e(B_{1}),t+1)=1
\end{eqnarray}
which implies that all $e(B_{1})$'s correspond to the blank from
cell $-t$ to cell $t$ in the simulated quantum Turing machine
$M_{1}$;
\begin{eqnarray}
&&\delta_{2}(q_{0}^{'},\underbrace{e(B_{1}),\ldots,e(B_{1}),n_{1}}_{t+1},
n_{2},\ldots,n_{l},e(B_{1}),\ldots,e(B_{1}),t+1,B_{3},\\
&&\hskip 10mm
p_{0},\underbrace{e(B_{1}),\ldots,e(B_{1}),n_{1}}_{t+1},
n_{2},\ldots,n_{l},e(B_{1}),\ldots,e(B_{1}),t+1)=1;
\end{eqnarray}
for $t+1\geq j\geq 1$,
\begin{eqnarray}
&&\nonumber\delta_{2}(p_{0},\underbrace{e(B_{1}),\ldots,e(B_{1}),n_{1}}_{t+1},
n_{2},\ldots,n_{l},e(B_{1}),\ldots,e(B_{1}),j,B_{3},\\
&&\hskip 10mm
p_{0},\underbrace{e(B_{1}),\ldots,e(B_{1}),n_{1}}_{t+1},
n_{2},\ldots,n_{l},e(B_{1}),\ldots,e(B_{1}),j-1)=1
\end{eqnarray}
and
\begin{eqnarray}
&&\nonumber\delta_{2}(p_{0},\underbrace{e(B_{1}),\ldots,e(B_{1}),n_{1}}_{t+1},
n_{2},\ldots,n_{l},e(B_{1}),\ldots,e(B_{1}),0,B_{3},\\
&&\hskip 10mm
q_{0},\underbrace{0,\ldots,0,n_{1}}_{t+1},n_{2},\ldots,n_{l},0,\ldots,0)=1
\end{eqnarray}
where $p_{0}$ is the initial state in $M_{1}$.

(ii) If $\delta_{1}(p_{0},\sigma_{1},\sigma,p_{1},d)=c$, then when
$d=R$,
\begin{eqnarray}
&&\nonumber\delta_{2}(p_{0},\underbrace{e(B_{1}),\ldots,e(B_{1}),
e(\sigma_{1})}_{t+1},\ldots,e(\sigma_{k}),e(B_{1}),\ldots,e(B_{1}),t+1,B_{4},\\
&&\nonumber\hskip 10mm
p_{1},\underbrace{e(B_{1}),\ldots,e(B_{1}),e(\sigma)}_{t+1},e(\sigma_{2}),\ldots,e(\sigma_{k}),e(B_{1}),\ldots,e(B_{1}),t+2)\\
&=&c;
\end{eqnarray}
 when $d=L$,
\begin{eqnarray}
&&\nonumber\delta_{2}(p_{0},\underbrace{e(B_{1}),\ldots,e(B_{1}),
e(\sigma_{1})}_{t+1},\ldots,e(\sigma_{k}),e(B_{1}),\ldots,e(B_{1}),t+1,B_{4},\\
&&\nonumber\hskip 10mm
p_{1},\underbrace{e(B_{1}),\ldots,e(B_{1}),e(\sigma)}_{t+1},e(\sigma_{2}),\ldots,e(\sigma_{k}),e(B_{1}),\ldots,e(B_{1}),t)\\
&=&c.
\end{eqnarray}

(iii) If $\delta_{1}(p,\sigma,\tau,q,d)=c$, then for any $1\leq
n_{j}\leq r-1$, $j=1,2,\ldots,i-1,i+1,\ldots,2t+1$, when $d=R$,
\begin{eqnarray}
&&\nonumber\delta_{2}(p,n_{1},n_{2},\ldots,n_{i-1},e(\sigma),n_{i+1},\ldots,n_{2t+1},i,B_{4},\\
&&\hskip 20mm
q,n_{1},n_{2},\ldots,n_{i-1},e(\tau),n_{i+1},\ldots,n_{2t+1},i+1)=c;
\end{eqnarray}
when $d=L$,
\begin{eqnarray}
&&\nonumber\delta_{2}(p,n_{1},n_{2},\ldots,n_{i-1},e(\sigma),n_{i+1},\ldots,n_{2t+1},i,B_{4},\\
&&\hskip 20mm
q,n_{1},n_{2},\ldots,n_{i-1},e(\tau),n_{i+1},\ldots,n_{2t+1},i-1)=c.
\end{eqnarray}

(iv) For any $p\in Q_{2}$, and any $j\in {\bf N}$,
\begin{equation}
\delta_{2}(p,n_{1},\ldots, n_{2t},j,\$,
p,n_{1},\ldots,n_{2t},j)=1,
\end{equation}
which means that after reading endmarker $\$$ a computation ends.

Next we will show that $\delta_{2}$ satisfies the well-formedness
conditions. Denote $C_{M_{1}}^{(b)}$, $C_{M_{2}}^{(b)}$, and
$C_{M_{2}}^{(b,+)}$ by

$C_{M_{1}}^{(b)}=\{|q\rangle|\tau\rangle|i\rangle: q\in Q_{1},
i\in [-t,t]_{{\bf Z}}, \tau\in
(\Sigma_{1}\cup\{B_{1}\})^{[-t,t]_{{\bf Z}}}\}$,

$C_{M_{2}}^{(b)}=\{|q\rangle|n_{1},\ldots,n_{2t+1}\rangle|n_{2t+2}\rangle:q\in
Q_{2}, 0\leq n_{i}\leq r-1, i=1,2,\ldots,2t+1, 1\leq n_{2t+2}\leq
2t+1\}$,

$C_{M_{2}}^{(b,+)}=\{|q\rangle|n_{1},\ldots,n_{2t+1}\rangle|n_{2t+2}\rangle:q\in
Q_{2}, 1\leq n_{i}\leq r-1, i=1,2,\ldots,2t+1, 1\leq n_{2t+2}\leq
2t+1\}$.

Since we assume that QTM $M_{1}$ will end within $t$ steps for any
input string $x$ with $|x|\leq n$, the evolution operator
$U_{M_{1}}$ restricted on $l_{2}(C_{M_{1}}^{(b)})$ is as: for any
$|q\rangle|\tau\rangle|i\rangle\in C_{M_{1}}^{(b)}$,
\begin{equation}
U_{M_{1}}|q\rangle|\tau\rangle|i\rangle=\sum_{p,\sigma,d\in\{-1,1\}}\delta_{1}(q,\tau(i),\sigma,p,d)|p\rangle|\tau(i,\sigma)\rangle|i+d\rangle
\end{equation}
where $d=-1$ and $d=1$ are identified with $d=L$ and $d=R$,
respectively, $\tau(i,\sigma)(j)=\left\{\begin{array}{ll} \sigma,&
{\rm if}\hskip 2mm j=i,\\
\tau(j),& {\rm otherwise},
\end{array}
\right.$ and $U_{M_{1}}$ is exactly unitary on subspace
$l_{2}(C_{M_{1}}^{(b)})$ of $l_{2}(C_{M_{1}})$, where $C_{M_{1}}$
is the set of all configurations of QTM $M_{1}$. Note that if
$|q\rangle|\tau\rangle|i\rangle\in C_{M_{1}}^{(b)}$, then we view
$\tau(j)=B_{1}$ for $|j|>t$. We now show that $\delta_{2}$
satisfies the W-F conditions Eqs. (9,10) by means of further
extensions. We consider them by dividing the following cases.

(1) {\it For the endmarker $\#$, $\delta_{2}$ satisfies the W-F
conditions}. For any $|q\rangle|n_{1},\ldots,n_{2t+1}\rangle\in
C_{M_{2}}^{(b)}$, if
$(q,n_{1},\ldots,n_{2t+2})\not=(q_{0},0,\ldots,0)$ or
$(q_{0},0,\ldots,0,1)$, then we define $\delta_{2}$ satisfies the
W-F conditions.
\begin{equation}
\delta_{2}(q,n_{1},\ldots,n_{2t+2},\#,q,n_{1},\ldots,n_{2t+2})=1;
\end{equation}
otherwise,
\begin{equation}
\delta_{2}(q,n_{1},\ldots,n_{2t+2},\#,q^{'},n_{1}^{'},\ldots,n_{2t+2}^{'})=1.
\end{equation}
Then by combining the definitions Eqs.(28,29) of 1) above with (1)
it is easy to check that for input $\#$, $\delta_{2}$ satisfies
the W-F conditions Eqs. (9,10).

(2) {\it For any input symbol $\sigma\in\Sigma_{1}$, $\delta_{2}$
satisfies the W-F conditions}. For any
$|q\rangle|n_{1},\ldots,n_{2t+2}\rangle \in C_{M_{2}}^{(b)}$, if
$(q,n_{1},n_{2},\ldots,n_{2t+2})\not=(q_{0},n_{1}^{'},\ldots,n_{i-1}^{'},0,\ldots,0,i)$
or $(q_{0},0,\ldots,0,j)$ for any $1\leq i\leq 2t+1$ and $1\leq
j\leq 2t+2$ with $n_{j}^{'}>0$, $j=1,2,\ldots,i-1$, then we define
\begin{equation}
\delta_{2}(q,n_{1},n_{2},\ldots,n_{2t+2},\sigma,q,n_{1},n_{2},\ldots,n_{2t+2})=1;
\end{equation}
and for the other cases we define
\begin{equation}
\delta_{2}(q,n_{1},n_{2},\ldots,n_{2t+2},\sigma,q^{'},n_{1}^{'},n_{2}^{'},\ldots,n_{2t+2}^{'})=0.
\end{equation}
As well, it is ready to check that for  any input
$\sigma\in\Sigma_{1}$, $\delta_{2}$ satisfies the W-F conditions
Eqs. (9,10) in $C_{M_{2}}^{(b)}$, and therefore,
$V_{\sigma}^{M_{2}}$ defined by Eq. (7) is a unitary operator on
$l_{2}(C_{M_{2}}^{(b)})$.

(3) {\it For tape symbol $B_{2}$, $\delta_{2}$ satisfies the W-F
conditions}. For any $|q\rangle|n_{1},\ldots,n_{2t+2}\rangle\in
C_{M_{2}}^{(b)}$, if

$(q,n_{1},n_{2},\ldots,n_{2t+2})\not=(q_{0},n_{1}^{'},n_{2}^{'},\ldots,n_{l}^{'},0,\ldots,0,i)$
or $(q_{0},\underbrace{0,\ldots,0,n_{1}},\ldots,n_{l},0,\ldots,s)$
for any $0\leq l\leq n$ and $0\leq s\leq l+1$ with $0\leq i\leq
l+1$ and $n_{j},n_{j}^{'}>0$, $j=1,2,\ldots,l$, then we define
\begin{equation}
\delta_{2}(q,n_{1},n_{2},\ldots,n_{2t+2},B_{2},q,n_{1},n_{2},\ldots,n_{2t+2})=1,
\end{equation}
and for the other cases we define
\begin{equation}
\delta_{2}(q,n_{1},n_{2},\ldots,n_{2t+2},B_{2},q^{'},n_{1}^{'},n_{2}^{'},\ldots,n_{2t+2}^{'})=0.
\end{equation}
Then by combining 3) and the above definitions of $\delta_{2}$ we
can easily know that for input $B_{2}$, $\delta_{2}$ satisfies the
W-F conditions on $C_{M_{2}}^{(b)}$, and therefore,
$V_{B_{2}}^{M_{2}}$ defined by Eq. (7) is a unitary operator on
$l_{2}(C_{M_{2}}^{(b)})$.

(4) {\it For tape symbol $B_{3}$, $\delta_{2}$ satisfies the W-F
conditions}. For any $|q\rangle|n_{1},\ldots,n_{2t+2}\rangle\in
C_{M_{2}}^{(b)}$, if

$(q,n_{1},n_{2},\ldots,n_{2t+2})\not=(q_{0},\underbrace{0,\ldots,0,n_{1}^{'}}_{t+1},n_{2}^{'},\ldots,n_{l}^{'},0,\ldots,0,i)$, or\\
$(p_{0},\underbrace{e(B_{1}),\ldots,e(B_{1}),n_{1}}_{t+1},n_{2},\ldots,n_{l},e(B_{1}),\ldots,e(B_{1}),j)$ with $0\leq j\leq t+1$,\\
or
$(q_{0}^{'},\underbrace{e(B_{1}),\ldots,e(B_{1}),n_{1}^{'}}_{t+1},\ldots,n_{l}^{'},e(B_{1}),\ldots,e(B_{1}),t+1)$\\
for any $0<l\leq n$, any $0<n_{j}^{'}<r-1$, $j=1,2,\ldots,l$, and
$t+1\geq i\geq 0$, then we define
\begin{equation}
\delta_{2}(q,n_{1},n_{2},\ldots,n_{2t+2},B_{3},q,n_{1},n_{2},\ldots,n_{2t+2})=1,
\end{equation}
and for the other cases we define $\delta_{2}=0$ for input
$B_{3}$.

Similar to the discussion of (3) above, we know that
$V_{B_{3}}^{M_{2}}$ defined by Eq. (7) is unitary on
$l_{2}(C_{M_{2}}^{(b)})$.

(5) Finally, we consider the case of input $B_{4}$. If mapping

$g:C_{M_{1}}^{(b)}\rightarrow C_{M_{2}}^{(b,+)}$\\
is defined by

$g|q\rangle|\tau\rangle|i\rangle=|q\rangle|e(\tau(-t)),e(\tau(-t+1)),\ldots,e(\tau(t))\rangle|i+t+1\rangle,$\\
then $g$ is a bijective mapping from $C_{M_{1}}^{(b)}$ to
$C_{M_{2}}^{(b,+)}$. By means of the definition $\delta_{2}$ for
input $B_{4}$, we know that the unitarity of $U_{M_{1}}$ on
$l_{2}(C_{M_{1}}^{(b)})$ implies that $V_{B_{4}}$ is also a
unitary operator on $l_{2}(C_{M_{1}}^{(b,+)})$. Indeed, if
\begin{equation}
U_{M_{1}}|q\rangle|\tau\rangle|i\rangle=\sum_{p\in Q_{1},\sigma\in
\Sigma_{1},d\in\{-1,1\}}\delta_{1}(q,\tau(i),\sigma,p,d)|p\rangle|\tau_{i}^{\sigma}\rangle|i+d\rangle,
\end{equation}
then
\begin{eqnarray}
V_{B_{4}}^{M_{2}}g(|q\rangle|\tau\rangle|i\rangle)&=&\nonumber
V_{B_{4}}^{M_{2}}|q\rangle|e(\tau(-t)),e(\tau(-t+1)),\ldots,e(\tau(i)),\ldots,e(\tau(t))\rangle|i+t+1\rangle\\
&=&\nonumber \sum_{p\in
Q_{1},\sigma\in\Sigma_{1},d\in\{-1,1\}}\delta_{2}(q,e(\tau(-t)),e(\tau(-t+1)),\ldots,e(\tau(i)),\ldots,e(\tau(t)),\\
&&\nonumber \hskip 10mm B_{4},p,
e(\tau(-t)),e(\tau(-t+1)),\ldots,e(\sigma),\ldots,e(\tau(t))\\
&&\nonumber \hskip 10mm |p\rangle|e(\tau(-t)),e(\tau(-t+1)),\ldots,e(\sigma),\ldots,e(\tau(t))\rangle\\
&=&\sum_{p\in
Q_{1},\sigma\in\Sigma_{1},d\in\{-1,1\}}\delta_{1}(q,\tau(i),\sigma,p,d)
|p\rangle g(|p\rangle|\tau_{i}^{\sigma}\rangle|i+d\rangle).
\end{eqnarray}
Therefore, $V_{B_{4}}^{M_{2}}$ is unitary on
$l_{2}(C_{M_{1}}^{(b,+)})$. Of course, $V_{B_{4}}^{M_{2}}$ can be
extended to be a unitary operator on $l_{2}(C_{M_{1}}^{(b)})$.

The rest is to see that $P_{a}^{M_{1}}(x)=P_{a}^{M_{2}}(x)$ for
any input string $x\in \Sigma_{1}^{*}$. It exactly follows from
the above definitions regarding $\delta_{2}$, and therefore, this
completes the proof. $\Box$

\end{document}